\documentclass[conference]{IEEEtran}
\usepackage{balance}
\usepackage[numbers,sort&compress]{natbib}
\usepackage{graphicx}
\graphicspath{{figs/}{figs/}} 
\DeclareGraphicsExtensions{.eps,.pdf}
\usepackage{amsmath}
\usepackage{amssymb}
\usepackage{subcaption}
\usepackage{caption}
\usepackage{multirow,multicol}
\usepackage{lipsum}
\usepackage{capt-of}

\begin{document}
	\title{Impact of Grid Impedance Variations on Harmonic Emission of Grid-Connected Inverters}
	\author{\IEEEauthorblockN{Bakhtyar Hoseinzadeh, Claus Leth Bak}
		\IEEEauthorblockA{Aalborg University, Denmark\\
			\{bho,clb\}@et.aau.dk}
	}
	\maketitle

\begin{abstract}
This paper addresses harmonic magnification due to resonance circuits resulting from interaction between uncertain grid impedance and converter. The source of harmonic may be either the grid or inverter. It is demonstrated that unknown and unpredictable grid impedance may result in variable resonance frequency, which challenges robust design of LCL filter of inverter.
\end{abstract}


\section{Introduction}\label{sec:1}
Nowadays, the widespread energy demand and global environmental concerns have accelerated development of renewable energy sources, particularly wind and solar power \cite{114,1019,a1,a2,a5}. Rapid growth in penetration of renewable energy sources into power system by further installation of them or replacement of existing conventional power plants, gradually weakens dominancy of synchronous machines compared to the power electronic based energy sources \cite{110}. The built-in inverter of aforementioned non synchronous generations, significantly affects reliability and security margin of operation and control of power system \cite{113,108,a33,a44}.

Prevalence of high order harmonics, which is one of side effects of built-in power electronic converters in renewable energy sources, inevitably imposes sharp changes on voltage, current and frequency profiles \cite{109}. As a result, the harmonic level at point of common coupling may violate the permissible limits recommended by relevant standards and grid codes \cite{112}.

The LCL filters employed in grid connected inverters are used to pass the fundamental frequency and attenuate the rest of undesired high order harmonics, which are appeared in the current and voltage profiles. Although, this goal is more or less achieved as an advantage of LCL filters, they interact with the grid elements as a part of a resonance circuit at a point close to the source of harmonics, i.e. inverter \cite{1020}. Harmonic emission of inverter or the harmonics, which are coming form the grid side may trigger existing parallel and/or series resonance circuits. This phenomenon may magnify harmonic level exceeding the limits recommended by grid codes and standards leading to distorted and hence undesired waveforms or in severe cases instability of inverter \cite{1000}.

In power electronic design, the LCL filter is designed in such a way that the resonance frequency of LCL is located far from the fundamental frequency \cite{1021}. In such a design procedure, the grid impedance variation are not considered in design procedure due high uncertainty in prediction of grid impedance, while the change in grid impedance results in remarkable shift in resonance frequency \cite{1022}.

Variable and unpredictable grid impedance affects the resonance frequency as an external factor, while there is an internal factor, which is also not considered in the LCL filter design \cite{1019,1021}. The scale of wind park, i.e. number of parallel/radial connection of wind park brunches changes the resonance frequency. It means that increment in number of brunches, directly but not linearly decreases the resonance frequency.

Estimation of grid dynamical behavior is rigorous due to its time-variant topology, diversity of energy sources, unknown and unpredictable demand response. Stochastic nature and intermittent connectivity of high share of renewable energy sources further deteriorate the situation \cite{103}. It means that it was assumed that the grid just consists of conventional power system elements, e.g. synchronous machines, transmission lines and power electronic-free loads. Nowadays, not only there are plenty of inverter-based generation sources behind the grid, but also the loads even contain harmonic emitter power electronic devices. Therefore, modeling the grid by a simple voltage source with only fundamental frequency in series with a set of constant inductor and resistor circuit may not lead to a comprehensive evaluation of any recently proposed scientific solution dealing with current status of grid behavior \cite{111}.
\section{Frequency Sweep of Grid Impedance}\label{sec:2}
\subsection{Field Measurement of Grid Impedance}\label{sec:3}
\begin{figure}[t]
\centering
\includegraphics[width=1\linewidth]{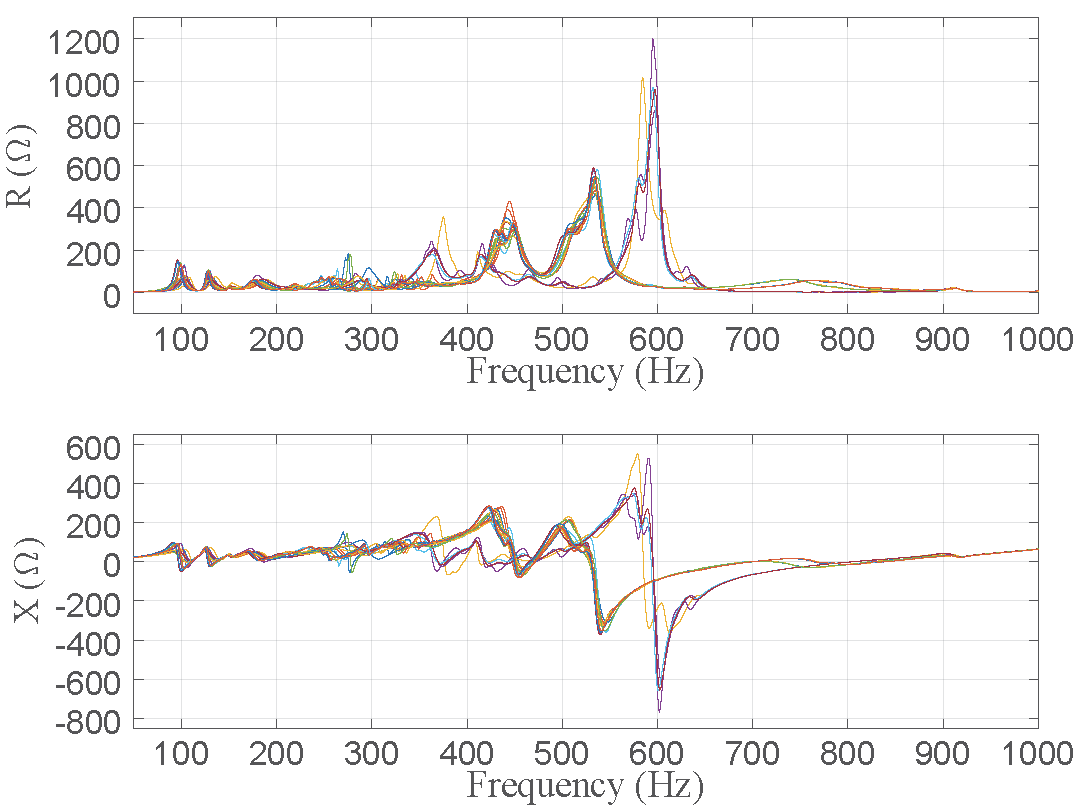}
\caption{24 frequency sweep snapshots of grid impedance}
\label{fig:1}	
\end{figure}
A set of field measurements have been fulfilled to carry out frequency sweep of grid impedance at point of common coupling in a wind park. The offshore wind park consists of a large number of wind turbines interconnected by collection network. The configuration of collection network includes several radial branches in parallel to harvest the power from wind turbines. In each branch, a set of wind turbines are connected together in parallel by short submarine cables.
\subsection{Frequency Sweep Analysis}\label{sec:4}
Fig.~\ref{fig:1} indicates 24 distinct snapshots of grid impedance measured at point of common coupling using frequency sweep through the range from 0 to 1 KHz. The field measurement are carried out in different times under normal operation conditions. Various characteristics/behaviors in frequency response of the grid impedance are illustrated in this section. A practical/realistic range for grid impedance variations under different conditions of grid is investigated, which should be considered to robust design of inverter controller and filter against grid impedance uncertainty.

The data are originally provided in Cartesian coordinates format, i.e. $ R(\omega)+jX(\omega) $, while its study in Polar coordinates ($ |Z|\angle Z$) is more common. Although, available snapshot data may not reflect precise and general overview of grid impedance, their remarkable correlation, consistency and accordance in most of cases is a good sign for analysis. It means that the disparity of data is not widespread and their classification seems possible to reach a solid conclusion about frequency response of wind farm collection network. However, the grid spectrum demonstrated in fig.~\ref{fig:1} can be analyzed from different aspect of view:

First, the frequency sweep clearly reveals that the grid reactance unexpectedly becomes negative per some particular frequency ranges, which means that modeling of grid equivalent impedance using simple and popular assumption of $ R+jL\omega $ may not be a precise hypothesis for a frequency range far from fundamental, i.e. for harmonic studies.

Second, a glance observation confirms that considering a particular and/or limited range for grid impedance may not be realistic in practice. More important, the percentage of variations (uncertainty) is out of engineering imagination (e.g. almost 20 times around 600 Hz), which may challenge any existing inverter control scheme.

Third, the dependency of grid reactance to frequency is expected, since the reactance of grid elements, e.g. cables/transformers, is inherently frequency dependent \cite{1020}. Fig.~\ref{fig:1} shows dependency of grid resistance to frequency, which should be investigated. The resistance of each individual element, itself, is not frequency dependent or its corresponding sensitivity is negligible at least for frequency range of interest in harmonic study. The grid equivalent resistance may vary in term of frequency, if a part of it is bypassed by an internal LC resonant circuit at a particular frequency range as it is utilized to bypass the damping resistor at switching frequency in C-type LCL filter design.

\begin{figure}[t]
	\begin{subfigure}[t]{0.5\linewidth}
		\centering
		\includegraphics[width=1.0\linewidth]{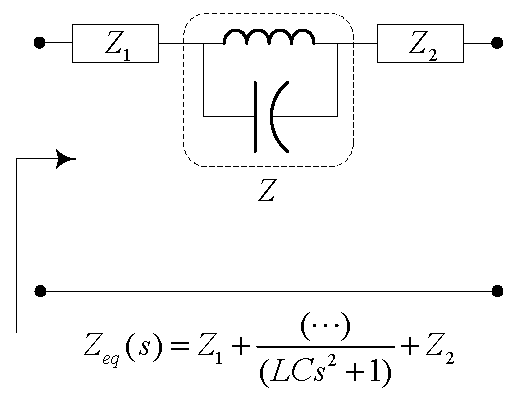}
		\caption{Parallel resonance in series}
		\label{fig:2}
	\end{subfigure}%
	\begin{subfigure}[t]{0.5\linewidth}
		\centering
		\includegraphics[width=1.0\linewidth]{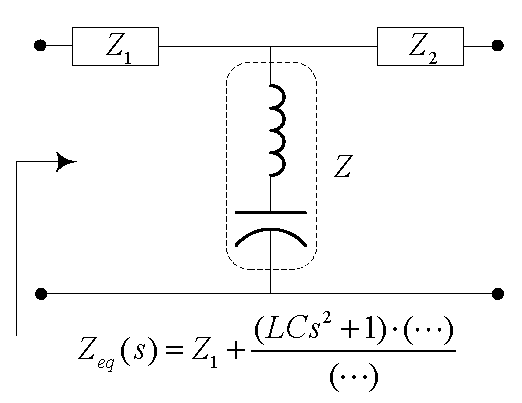}
		\caption{Series resonance in parallel}
		\label{fig:3}
	\end{subfigure}%
	\caption{Zeros in impedance transfer function}
	\label{fig:4}
\end{figure}

Sharp plunges are observable in both resistance and reactance profiles, e.g. around 100, 450 and 600 Hz, which may be due to existence of nearby/distant series and/or parallel resonance circuits. Fig.~\ref{fig:4} indicates different possibility of resonant circuits, which may come to exist due to interaction between grid impedance, inverter LCL filter and inverter output impedance. Fig.~\ref{fig:2} shows the parallel resonance circuit in series with grid and inverter impedance. If the frequency of a given current harmonic produced by either grid or inverter accidentally matches the frequency of parallel resonance circuit, the limited/negligible current harmonic is converted to a considerable voltage harmonic by passing through impedance $z$, which is theoretically infinite and practically a large value. Similar phenomenon may happen for dual circuit depicted in Fig.~\ref{fig:3}.

High/low amplitude of $z$ in Fig.~\ref{fig:2}/\ref{fig:3} may lead to magnification of voltage/current harmonics, which thereafter can be also appeared in current/voltage profile by passing through the rest of existing elements, i.e. $z_1$ and $z_2$. This phenomenon may be frequently repeated in different frequencies, which is the main concern of this paper.

There is an particular curve in Fig.~\ref{fig:1}, which is inconsistent with the others (yellow). This data has been measured at the same bus, but in different period of time and operation condition of both grid and wind power plant. However, it does not properly follow the overall trend of other patterns in term of behavior and therefore it enervates/challenges the aforementioned agreement about frequency response of grid impedance.
\section{Simulation Setup}\label{sec:6}
\begin{figure}[t]
	\centering
	\includegraphics[width=1.0\linewidth]{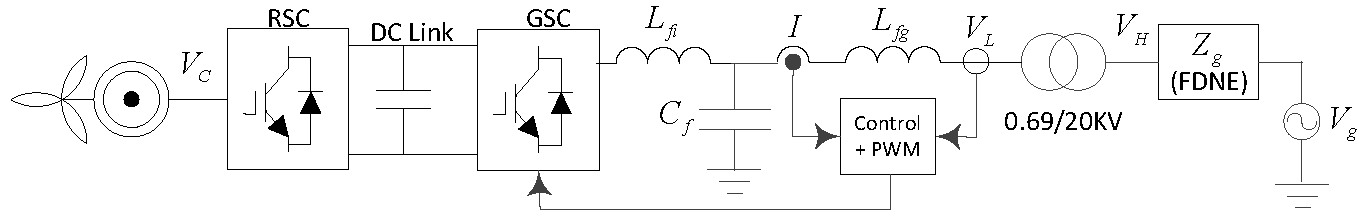}
	\caption{Overall layout of grid-connected wind turbine}
	\label{fig:5}
\end{figure}
In order to investigate the impact of grid impedance variations on performance and stability of grid connected wind turbine, simulations are carried out in PSCAD 4.6 software. Fig.~\ref{fig:5} indicates the structure of studied wind turbine with focus on inner control loop of inverter. Each individual data of measured grid impedance cases are converted to corresponding impedance transfer function using a built-in block called Frequency Dependant Network Equivalent (FDNE) in PSCAD.
\section{Simulation Results}\label{sec:7}

\begin{figure}[t]
	\centering
	\includegraphics[width=1.0\linewidth]{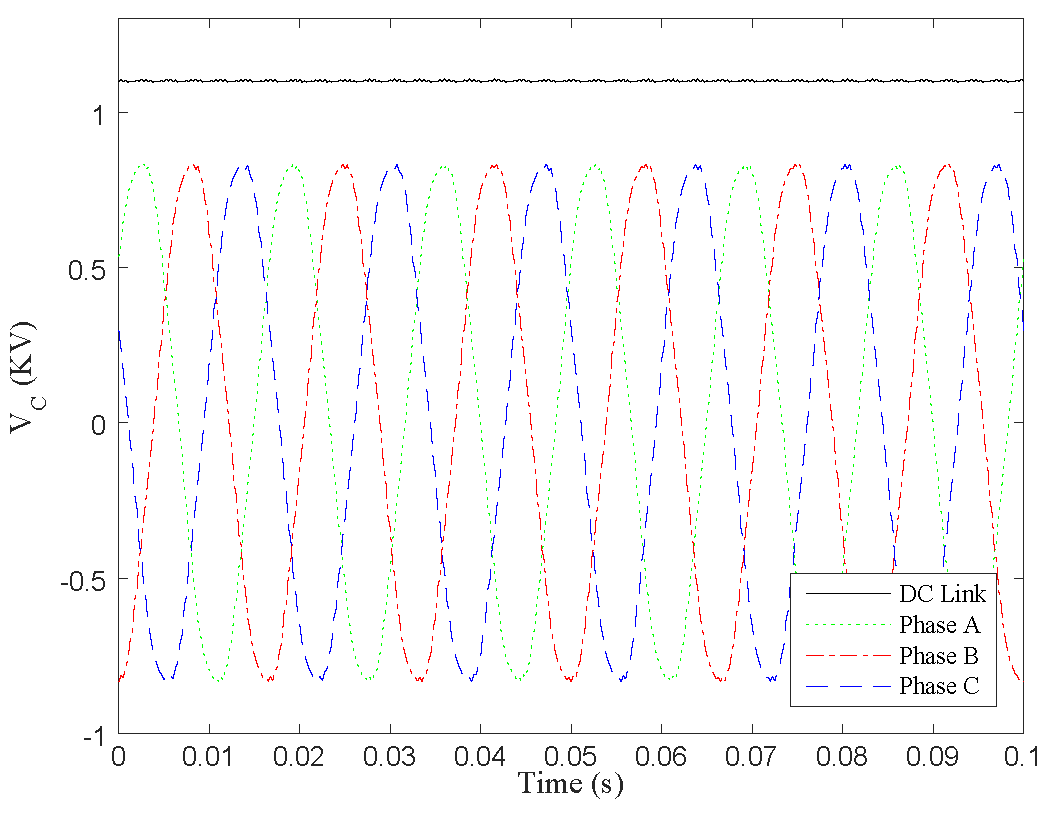}
	\caption{DC Link \& generator voltage}
	\label{fig:6}
\end{figure}
\begin{figure}[t]
	\centering
	\includegraphics[width=1.0\linewidth]{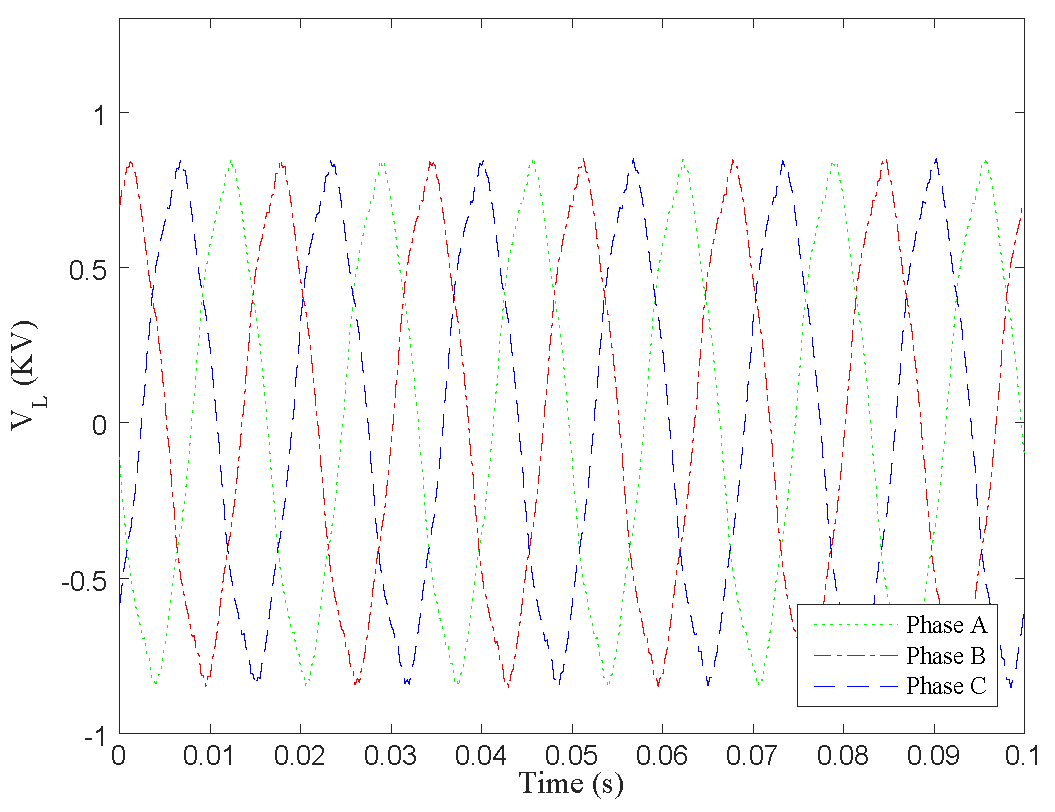}
	\caption{Inverter voltage}
	\label{fig:7}
\end{figure}
\begin{figure}[t]
	\centering
	\includegraphics[width=1.0\linewidth]{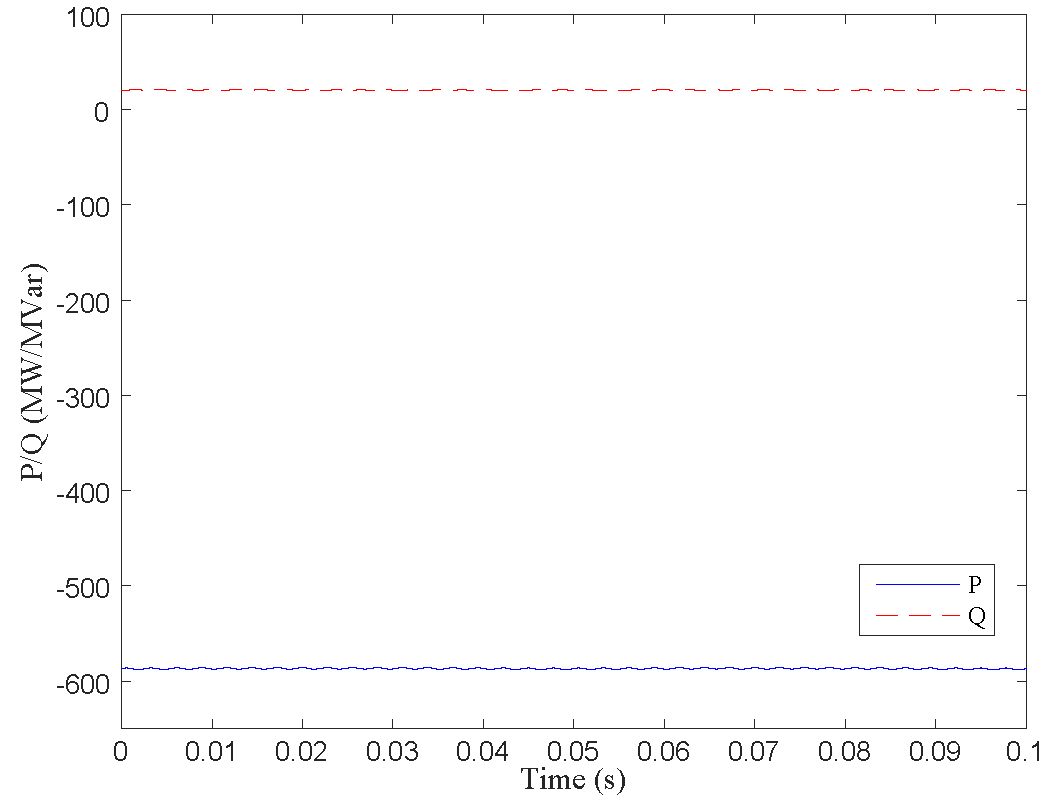}
	\caption{Active \& reactive power of inverter}
	\label{fig:8}
\end{figure}
\begin{figure}[t]
	\centering
	\includegraphics[width=1.0\linewidth]{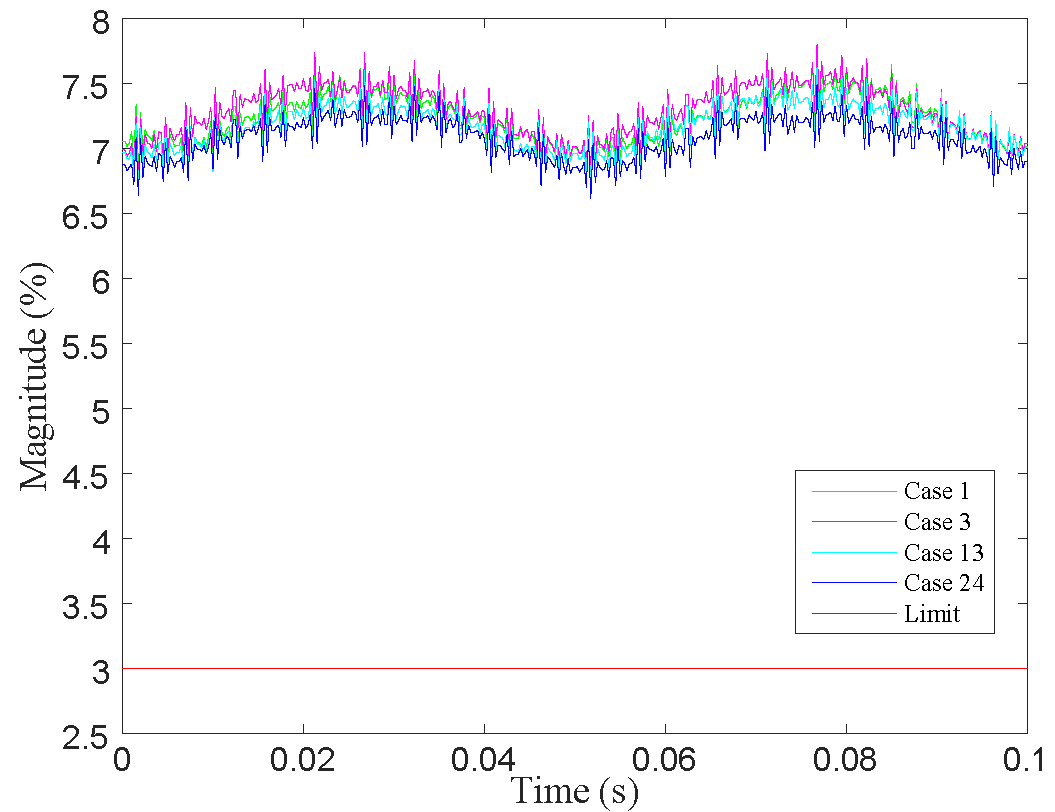}
	\caption{Third harmonic amplitude}
	\label{fig:9}
\end{figure}

Figs.~\ref{fig:6}-\ref{fig:9} indicate the simulation results in steady state throughout 0.1 second of time. The waveforms depicted in Fig.~\ref{fig:6} are the voltage of DC link and three phase voltages of generator ($ V_C $) as input of converter, respectively. The number of minor peaks in DC link voltage profile per one period of converter waveform is equal to the number of power electronic switches, i.e. 6, which is explicitly observable. Fig.~\ref{fig:7} shows the output voltage of inverter ($ V_L $). Comparing the voltage waveforms of converter and inverter clearly indicates that the output of inverter is relatively distorted and polluted by harmonics. The amplitude of voltage curves in both figures are almost equal to 850 V or equal to 600 V in $ rms $ quantity.

Fig.~\ref{fig:8} indicates active (solid blue line) and reactive (dashed red line) power of inverter equal to -0.58 MW and 20 MVar, respectively. Due to dominancy of fundamental harmonic comparing to the other high order harmonics, the simulation results associated with the fundamental harmonic of all cases 1, 3, 13 and 24 are almost similar to each other. Therefore, for the sake of similarity and space limitation in the paper, only the results of case 24 have been presented.

The third harmonic magnitude of cases 1, 3, 13 and 24 are plotted in Fig.~\ref{fig:9} with different colors. The permissible limit of third harmonic recommended by standard \cite{1018} is equal to 3 \%, which has been specified with a solid and flat red color line. As can be seen, the aforementioned limit is violated by a high quantity of harmonic equal to almost 7\%. It means that the third harmonic is magnified by resonance circuits coming from interaction of grid impedance and inverter internal impedance.
\section{Conclusion}
Collection network of wind farms including lines, transformers and LCL filter of inverters in connection with grid impedance may constitute a LC resonance circuit. Grid impedance variation and uncertainty changes the resonance frequency of resulting LC circuits in a widespread range. The resonance circuit/s may magnify harmonics produced by inverters or the harmonics coming from the grid side.
\balance
\bibliographystyle{IEEEtran}
\bibliography{bibliography}

\begin{thebibliography}{19}
\providecommand{\natexlab}[1]{#1}
\providecommand{\url}[1]{\texttt{#1}}
\expandafter\ifx\csname urlstyle\endcsname\relax
  \providecommand{\doi}[1]{doi: #1}\else
  \providecommand{\doi}{doi: \begingroup \urlstyle{rm}\Url}\fi

\bibitem[Hoseinzadeh et~al.(2015{\natexlab{a}})Hoseinzadeh, Da~Silva, and
  Bak]{114}
Bakhtyar Hoseinzadeh, Filipe~Faria Da~Silva, and Claus~Leth Bak.
\newblock Improved lvrt grid code under islanding condition.
\newblock In \emph{Industrial Electronics Society, IECON 2015-41st Annual
  Conference of the IEEE}, pages 000421--000426. IEEE, 2015{\natexlab{a}}.

\bibitem[Pan et~al.(2014)Pan, Ruan, Bao, Li, and Wang]{1019}
Donghua Pan, Xinbo Ruan, Chenlei Bao, Weiwei Li, and Xuehua Wang.
\newblock Capacitor-current-feedback active damping with reduced computation
  delay for improving robustness of lcl-type grid-connected inverter.
\newblock \emph{IEEE Transactions on Power Electronics}, 7\penalty0
  (29):\penalty0 3414--3427, 2014.

\bibitem[Zeinalzadeh et~al.(2016{\natexlab{a}})Zeinalzadeh, Ghorbani, and
  Yee]{a1}
Ashkan Zeinalzadeh, Reza Ghorbani, and James Yee.
\newblock Stochastic model of voltage variations in the presence of
  photovoltaic systems.
\newblock In \emph{American Control Conference (ACC), 2016}, pages 5032--5037.
  American Automatic Control Council (AACC), 2016{\natexlab{a}}.

\bibitem[Zeinalzadeh et~al.(2013)Zeinalzadeh, Ghorbani, and Reihani]{a2}
Ashkan Zeinalzadeh, Reza Ghorbani, and Ehsan Reihani.
\newblock Optimal power flow problem with energy storage, voltage and reactive
  power control.
\newblock The 45th ISCIE International Symposium on Stochastic Systems Theory
  and Its Applications, 2013.

\bibitem[Zeinalzadeh et~al.(2016{\natexlab{b}})Zeinalzadeh, Ghorbani, and
  Yee]{a5}
Ashkan Zeinalzadeh, Reza Ghorbani, and James Yee.
\newblock {Voltage control in the presence of photovoltaic systems}.
\newblock \emph{ArXiv e-prints}, December 2016{\natexlab{b}}.

\bibitem[Boroojeni et~al.(2016)Boroojeni, Amini, Nejadpak, Iyengar,
  Hoseinzadeh, and Bak]{110}
K.~G. Boroojeni, M.~H. Amini, A.~Nejadpak, S.~S. Iyengar, Bakhtyar Hoseinzadeh,
  and C.~L. Bak.
\newblock A theoretical bilevel control scheme for power networks with
  large-scale penetration of distributed renewable resources.
\newblock \emph{2016 IEEE International Conference on Electro Information
  Technology (EIT)}, pages 0510--0515, 2016.

\bibitem[Hoseinzadeh(2015)]{113}
Bakhtyar Hoseinzadeh.
\newblock A power system emergency control scheme in the presence of high wind
  power penetration.
\newblock \emph{Department of Energy Technology}, pages 1--123, 2015.

\bibitem[Kalogiannis et~al.(2015)Kalogiannis, Muller~Llano, Hoseinzadeh, and
  Silva]{108}
Theodoros Kalogiannis, Enrique Muller~Llano, Bakhtyar Hoseinzadeh, and
  Filipe~Faria Silva.
\newblock Impact of high level penetration of wind turbines on power system
  transient stability.
\newblock \emph{Power Tech Conference Proceedings, 2015 IEEE}, pages 1--6,
  2015.

\bibitem[{Zeinalzadeh} et~al.(2016){Zeinalzadeh}, {Chakraborty}, and
  {Gupta}]{a33}
A.~{Zeinalzadeh}, I.~{Chakraborty}, and V.~{Gupta}.
\newblock {Pricing energy in the presence of renewables}.
\newblock \emph{ArXiv e-prints}, November 2016.

\bibitem[{Zeinalzadeh} and {Gupta}(2016)]{a44}
A.~{Zeinalzadeh} and V.~{Gupta}.
\newblock {Minimizing risk of load shedding and renewable energy curtailment in
  a microgrid with energy storage}.
\newblock \emph{ArXiv e-prints}, November 2016.

\bibitem[Hoseinzadeh et~al.(2015{\natexlab{b}})Hoseinzadeh, Faria~Silva, and
  Leth~Bak]{109}
Bakhtyar Hoseinzadeh, Filipe Faria~Silva, and Claus Leth~Bak.
\newblock Decentralized coordination of load shedding and plant protection
  considering high share of ress.
\newblock \emph{Power Systems, IEEE Transactions on}, 2015{\natexlab{b}}.

\bibitem[Chaudhary et~al.(2016)Chaudhary, Lascu, Hoseinzadeh, Teodorescu,
  Kocewiak, Sørensen, and Jensen]{112}
Sanjay~Kumar Chaudhary, Cristian~Vaslie Lascu, Bakhtyar Hoseinzadeh, Remus
  Teodorescu, Lukasz Kocewiak, Troels Sørensen, and Christian~F. Jensen.
\newblock Challenges with harmonic compensation at a remote bus in offshore
  wind power plant.
\newblock \emph{IEEE International Conference on Environment and Electrical
  Engineering (EEEIC 2016)}, pages 1--5, 2016.

\bibitem[Li et~al.(2015)Li, Wu, Geng, Yuan, Xia, and Zhang]{1020}
Xiaoqiang Li, Xiaojie Wu, Yiwen Geng, Xibo Yuan, Chenyang Xia, and Xue Zhang.
\newblock Wide damping region for lcl-type grid-connected inverter with an
  improved capacitor-current-feedback method.
\newblock \emph{Power Electronics, IEEE Transactions on}, 30\penalty0
  (9):\penalty0 5247--5259, 2015.

\bibitem[Rocabert et~al.(2012)Rocabert, Luna, Blaabjerg, and Rodríguez]{1000}
Joan Rocabert, Alvaro Luna, Frede Blaabjerg, and Paul Rodríguez.
\newblock Control of power converters in ac microgrids.
\newblock \emph{Power Electronics, IEEE Transactions on}, 27\penalty0
  (11):\penalty0 4734--4749, 2012.

\bibitem[Bao et~al.(2014)Bao, Ruan, Wang, Li, Pan, and Weng]{1021}
Chenlei Bao, Xinbo Ruan, Xuehua Wang, Weiwei Li, Donghua Pan, and Kailei Weng.
\newblock Step-by-step controller design for lcl-type grid-connected inverter
  with capacitor--current-feedback active-damping.
\newblock \emph{Power Electronics, IEEE Transactions on}, 29\penalty0
  (3):\penalty0 1239--1253, 2014.

\bibitem[Xu et~al.(2014)Xu, Xie, and Tang]{1022}
Jinming Xu, Shaojun Xie, and Ting Tang.
\newblock Active damping-based control for grid-connected-filtered inverter
  with injected grid current feedback only.
\newblock \emph{Industrial Electronics, IEEE Transactions on}, 61\penalty0
  (9):\penalty0 4746--4758, 2014.

\bibitem[Hoseinzadeh et~al.(2015{\natexlab{c}})Hoseinzadeh, Faria~Silva, and
  Leth~Bak]{103}
Bakhtyar Hoseinzadeh, Filipe Faria~Silva, and Claus Leth~Bak.
\newblock Adaptive tuning of frequency thresholds using voltage drop data in
  decentralized load shedding.
\newblock \emph{Power Systems, IEEE Transactions on}, 30\penalty0 (4):\penalty0
  2055--2062, July 2015{\natexlab{c}}.

\bibitem[Hoseinzadeh and Bak(2016)]{111}
Bakhtyar Hoseinzadeh and Claus~Leth Bak.
\newblock Admittance modeling of voltage and current controlled inverter for
  harmonic instability studies.
\newblock \emph{PES General Meeting Conference Exposition, 2016 IEEE}, pages
  1--5, 2016.

\bibitem[IEEE(2014)]{1018}
IEEE.
\newblock Ieee recommended practice and requirements for harmonic control in
  electric power systems.
\newblock \emph{IEEE Std 519-2014}, pages 1--29, June 2014.
\newblock \doi{10.1109/IEEESTD.2014.6826459}.

\end{thebibliography}
\end{document}